\documentclass[preprint,showpacs,preprintnumbers,amsmath,amssymb]{revtex4}

\usepackage{graphicx}
\usepackage{dcolumn}
\usepackage{bm}

\begin{document}

\preprint{APS/123-QED}

\title{Heavy Fermion behaviour in PrRh$_{2}$B$_{2}$C: An excitonic mass enhancement}
\author{V. K. Anand}
\email{vivekkranand@gmail.com}
\author{Z. Hossain}
\affiliation{ Department of Physics, Indian Institute of Technology, Kanpur 208016, India}
\author{G. Chen}
\author{M. Nicklas}
\author{C. Geibel}
\affiliation{Max Planck Institute for Chemical Physics of Solids, 01187 Dresden, Germany}
\date{\today}
\begin{abstract}

           We report magnetic and transport properties of a new quaternary borocarbide PrRh$_{2}$B$_{2}$C, based on magnetisation, resistivity and specific heat studies. This compound forms in LuNi$_{2}$B$_{2}$C-type tetragonal structure (space group {\it I4/mmm}) and does not exhibit magnetic ordering or superconductivity down to 0.5 K. The crystal field analysis of specific heat data suggests a singlet ground state in this compound. The high value of the Sommerfeld coefficient, $\gamma$ $\approx$ 250 mJ/mole~K$^{2}$, together with a singlet ground state suggests that the heavy fermion behaviour in PrRh$_{2}$B$_{2}$C results from the interaction of the conduction electrons with the low lying crystal field excitations. No signature of magnetic ordering or superconductivity is observed in PrRh$_{2}$B$_{2}$C under the application of pressure up to 23 kbar.

\end{abstract}

\pacs{71.27.+a, 74.70.Dd, 71.70.Ch, 72.10.Di}
\maketitle


\section*{Introduction}

          The heavy fermion systems have fascinated the condensed matter physicists with the exciting physics in the vicinity of quantum critical point. Most of the known heavy fermion systems belong to Ce-, Yb- or U-based intermetallic compounds. Recently discovered Pr-based heavy fermion superconductor PrOs$_{4}$Sb$_{12}$ \cite{1} presents many unusual phenomena, such as, two distinct superconducting phases, time reversal symmetry breaking, and point nodes at Fermi surface etc. \cite{1,2,3,4,5} In contrast to the numerous Ce-based magnetically mediated heavy fermion superconductors, PrOs$_{4}$Sb$_{12}$ is the only known Pr-based heavy fermion superconductor. Further, in contrast to the case of Ce-compounds where spin Kondo effect leads to the heavy fermion behaviour, in Pr-compounds like PrInAg$_{2}$ and PrFe$_{4}$P$_{12}$ it is the quadrupolar Kondo effect which is suggested to be responsible for heavy fermion behaviour. \cite{6,7}

          Another important mechanism for mass enhancement in Pr-compounds is the inelastic scattering of conduction electrons by the angular momentum associated with the crystal electric field (CEF) levels, referred as excitonic mass enhancement. The theory of excitonic mass enhancement was proposed by White and Fulde \cite{8} to explain the mass enhancement in elemental Pr itself and subsequently extended to rare earth systems with a CEF-split nonmagnetic singlet ground state. \cite{9} The true realization of heavy fermion behaviour due to crystal field excitatons was found in PrOs$_{4}$Sb$_{12}$ with a Sommerfeld coefficient $\gamma$ $\approx$ 350 mJ/mole~K$^{2}$. \cite{5} Very recently we have seen evidence of excitonic mass enhancement in Pr$_{2}$Rh$_{3}$Ge$_{5}$ leading to moderate heavy fermion behaviour. \cite{10} We present further evidence for excitonic mass enahancement in PrRh$_{2}$B$_{2}$C in this paper. In addition to providing one more example of Pr-based heavy fermion compound through our analysis of specific heat data we attempt to provide a qualitative explanation of the unusual route to the heavy fermion state in this compound. PrRh$_{2}$B$_{2}$C is a new member of the quaternary borocarbide family that gives rise to the hope to provide a new route to high temperature superconductivity in boron based compounds which provided a unique playground to investigate the interplay between superconductivity and magnetism. We present in this paper electrical resistivity, magnetic susceptibility and specific heat data of a novel Pr-based quaternary compound. The reproducibility of the results has been checked by similar studies on a second batch of sample.

\section*{Experimental}

          We prepared polycrystalline samples of PrRh$_{2}$B$_{2}$C and the nonmagnetic reference compound LaRh$_{2}$B$_{2}$C starting with high purity elements (99.99 \% or better) in stoichiometric composition by the standard arc melting on water cooled copper hearth. During the arc melting process samples were flipped and melted several times to improve the homogeneity. Arc melted samples were annealed for a week at 1200 $^{o}$C under dynamic vacuum. Samples were characterized by copper K$_\alpha$ X-ray diffraction and scanning electron microscopy. A SQUID magnetometer was used for magnetisation measurements. Heat capacity was measured by relaxation method in a physical property measurement system (PPMS, Quantum Design), and electrical resistivity was measured by the standard ac four probe techniques using the AC-Transport option of PPMS. Electrical resistivity measurements under hydrostatic pressure were carried out in a double-layer piston-cylinder type pressure cell for pressures up to 23 kbar. Silicone fluid served as pressure transmitting medium. The pressure was determined at low temperatures by monitoring the pressure induced shift of the superconducting transition temperature of lead. The narrow width of the transition confirmed the good hydrostatic pressure conditions inside the pressure cell.

\section*{Results and discussion}

          The powder X-ray diffraction (XRD) data of a polycrystalline sample of PrRh$_{2}$B$_{2}$C were analyzed by WINXPOW software and further refined by least squares Rietveld refinement method using the FULLPROF software (Fig. 1), the quality parameter $\chi^2$ being 2.82. PrRh$_{2}$B$_{2}$C forms in LuNi$_{2}$B$_{2}$C-type tetragonal structure (space group \textit{I4/mmm}) with lattice constants \textit{a} = 3.855 \AA, \textit{c} = 10.257 \AA ~and unit cell volume = 152.44 \AA$^3$.  The nonmagnetic reference compound LaRh$_{2}$B$_{2}$C also crystallises in the same tetragonal structure with lattice constants \textit{a} = 3.896 \AA, \textit{c} = 10.247 \AA ~and unit cell volume = 155.53 \AA$^3$, which are in fairly good agreement with the values reported. \cite{11} From XRD and scanning electron microscope (SEM) image we estimate impurity phase(s) to be less than 3\% of the main phase.

\begin{figure}
\noindent \includegraphics[width=8.50cm, keepaspectratio]{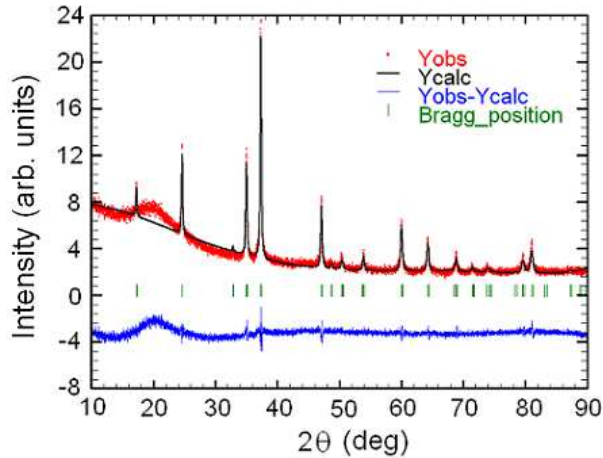}
\caption{\label{fig1} (colour online) Powder X-ray diffraction pattern of PrRh$_{2}$B$_{2}$C recorded at room temperature. The solid line through the experimental points is the Rietveld refinement profile calculated for LuNi$_{2}$B$_{2}$C-type tetragonal {\it I4/mmm} structural model. The lowermost curve represents the difference between the experimental and model results.}
\end{figure}

          The magnetic susceptibility data of PrRh$_{2}$B$_{2}$C is shown in Fig. 2 as a function of temperature. No anomaly is seen in the susceptibility down to 2 K. At higher temperatures the susceptibility curve follows a Curie-Weiss behaviour $\chi = C/(T-\theta_{p}$). From the linear fit of inverse susceptibility data at 1.0 T we obtained the effective moment $\mu_{eff}$ = 3.59 $\mu_{B}$, which is very close to the theoretically expected value of 3.58 $\mu_{B}$ for Pr$^{3+}$ ion. The paramagnetic Curie-Weiss temperature, $\theta_{p}$ = -3.9 K. The inset of Fig. 2 shows the magnetic field dependence of isothermal magnetisation, $M(B)$ of PrRh$_{2}$B$_{2}$C.  The isothermal magnetization at 2 K shows slight nonlinearity, most likely due to the crystal field effects (the kink in $M(B)$ near 2.7 T is an experimental artifact). The magnetisation does not reach the saturation value up to 5 T, it attains a value of 1.17 $\mu_{B}$ at 5 T.

\begin{figure}
\includegraphics[width=8.5cm, keepaspectratio]{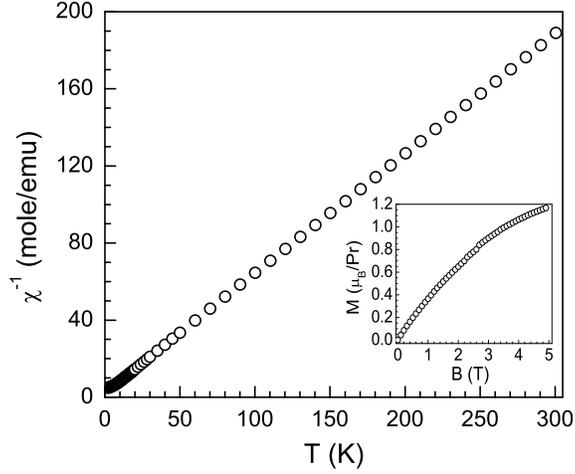}
\caption{\label{fig2} Temperature dependence of inverse magnetic susceptibility of PrRh$_{2}$B$_{2}$C measured in a field of 1.0 T. The inset shows the field dependence of isothermal magnetisation at 2 K.}
\end{figure}

\begin{figure}
\includegraphics[width=8.5cm, keepaspectratio]{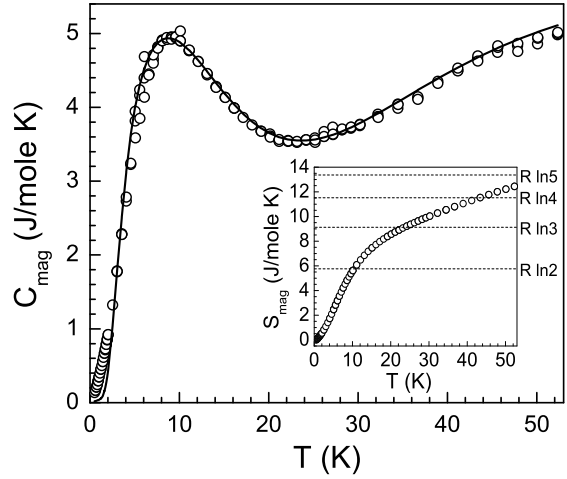}
\caption{\label{fig3} Magnetic part of specific heat of PrRh$_{2}$B$_{2}$C as a function of temperature. Solid line shows the fit to the CEF scheme as described in the text. Inset shows the temperature dependence of magnetic part of entropy.}
\end{figure}

          The specific heat data of PrRh$_{2}$B$_{2}$C do not show any anomaly corresponding to a magnetic or superconducting transition down to 0.5 K. However, a broad Schottky-type anomaly centered around 9 K is observed in the magnetic part of specific heat. The magnetic contribution to the specific heat, obtained by subtracting the specific heat of LaRh$_{2}$B$_{2}$C from that of PrRh$_{2}$B$_{2}$C, assuming the lattice contribution to be approximately equal to the specific heat of LaRh$_{2}$B$_{2}$C is shown in Fig. 3. The experimentally observed feature of the magnetic part of the specific heat above 2 K could be reproduced by a crystal electric field (CEF) analysis with four levels: three singlets at 0 K, 14 K, 36 K and a doublet at 155 K. The solid line in Fig. 3 represents the fit with this CEF level scheme. The inset of Fig. 3 shows the magnetic contribution to the entropy of PrRh$_{2}$B$_{2}$C. The magnetic entropy attains a value of $R ln(2)$ at 10 K supporting the proposed CEF level scheme of a singlet ground state lying below the first excited singlet at 14 K. From the specific heat data below 2 K we estimate a lower bound to the value of Sommerfeld coefficient $\gamma$ $\approx$ 250 mJ/mole K$^{2}$. It is observed that below 0.5 K $C/T$ increases with decreasing temperature (inset of Fig. 4) while $C$ keeps on decreasing. Such a behaviour of specific heat suggests a gradual onset of heavy fermion state and has some similarity with the $C/T$ upturn in heavy fermion compounds YbNi$_{2}$B$_{2}$C \cite{12} and YbRh$_{2}$Si$_{2}$. \cite{13} However, we can not entirely exclude the possibility that sharp increase of $C/T$ might be related to the onset of phase transition at further lower temperatures.

\begin{figure}
\includegraphics[width=8.5cm, keepaspectratio]{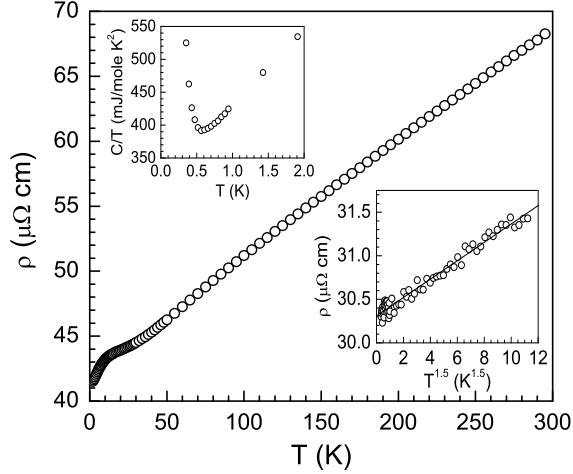}
\caption{\label{fig4} Temperature dependence of electrical resistivity of PrRh$_{2}$B$_{2}$C in the temperature range 0.5 K -- 300 K. The lower inset shows the electrical resistivity data below 5 K plotted as $\rho$ \textit{vs.} $T^{1.5}$ (measured on second sample). The solid line is a guide to eye. The upper inset shows the low temperature specific heat data below 2 K plotted as $C/T$ \textit{vs.} $T$.}
\end{figure}

            The electrical resistivity of PrRh$_{2}$B$_{2}$C shown in Fig. 4 exhibits metallic behaviour with $\rho_{300 K}$ = 68 $\mu \Omega$~cm, residual resistivity $\rho_{0}$ $\approx$ 42 $\mu \Omega$~cm and residual resistivity ratio RRR = $\rho_{300 K}$/$\rho_{0}$ $\approx$ 1.6. At higher temperatures the resistivity decreases almost linearly with decreasing temperature. We attribute the anomaly around 10 K to be due to crystal field effect which is consistent with the specific heat data. The electrical resistivity of LaRh$_{2}$B$_{2}$C which we measured down to 0.5 K also shows metallic behaviour. A rapid drop is observed in resistivity of the La-based compound below 3 K possibly due to incipient superconductivity.

          The low temperature electrical resistivity data of PrRh$_{2}$B$_{2}$C fits well with $\rho(T) = \rho_{0} + AT^{n}$ with n $\approx$ 1.5 and $A$ = 0.1 $\mu \Omega$~cm/K$^{n}$ and is shown in the inset of Fig. 4 plotted as $\rho$ \textit{vs.} $T^{1.5}$ below 5 K. A $T^{1.5}$ temperature dependence of the electrical resistivity is a characteristic of non-Fermi liquid behaviour as predicted by the spin fluctuation theory for a three dimensional system near an antiferromagnetic quantum critical point (AF-QCP). However in our compound because of the singlet ground state the departure from the Fermi liquid behaviour is more likely to be due to the presence of low-lying CEF levels. Further on, the residual resistivity is rather high which makes the power law exponent not so reliable. Therefore, we refrain from putting too much stress on the resistivity exponent and the associated non-Fermi liquid nature and proximity of PrRh$_{2}$B$_{2}$C to an AF-QCP.

          We also calculated the value of Wilson ratio using the value of $\chi$ = 0.25 emu/mole at 2 K and $\mu_{eff}$ = 3.59 $\mu_{B}$ together with $\gamma$ $\approx$ 250 mJ/mole~K$^{2}$ and obtained R$_{w}$ = 17 which is quite large. The high value of electronic specific heat coefficient $\gamma$ together with the enhanced value of R$_{w}$ is a clear indication of strong correlations and heavy fermion behaviour in this compound. Since no Kondo-like behaviour is observed in $\rho(T)$, the mechanism for the heavy fermion behaviour may be rooted in the low lying CEF splitting similar to that in the heavy fermion superconductor PrOs$_{4}$Sb$_{12}$. \cite{1,5} It was suggested by Fulde and Jensen \cite{9} that for a system with CEF-split singlet ground state and low CEF splitting energy the inelastic scattering of conduction electrons with the excited levels of angular momentum of the 4\textit{f} electrons can result in an enhanced mass of the conduction electrons. After a rigorous theoretical analysis they have shown that the enhanced mass due to the inelastic transition between two levels \textit{i} and \textit{j} can be found by the expression \cite{9},

\begin{displaymath}
 \frac{m*}{m} = 1 + (g_J - 1)^2 J_{sf}^2 N(0) \frac{2 \mid < i \mid J \mid j > \mid^2}{\Delta}
\end{displaymath}

\noindent where \textit{J$_{sf}$} is the exchange integral coupling between the conduction electrons and the \textit{f} electrons, and N(0) conduction electron density of states at the Fermi level. The matrix element between \textit{i} and \textit{j }can be calculated using the CEF parameters. An inelastic neutron scattering experiment on PrRh$_{2}$B$_{2}$C is highly desired to know the exact CEF level scheme and evaluate the matrix elements. The CEF analysis of specific heat data clearly suggests a CEF-split singlet ground state with another singlet excited state at about 14 K and therefore an excitonic mass enhancement by the crystal field excitations leading to the heavy fermion behaviour in PrRh$_{2}$B$_{2}$C.

Using similar CEF analysis of magnetic specific heat data we have seen that PrPd$_{2}$B$_{2}$C also has a singlet ground state lying below the first excited singlet at 50 K and a doublet at 80 K. \cite{14} However, the value of Sommerfeld coefficient $\gamma$ $\approx$ 16 mJ/mole~K$^{2}$ is much lower than that of PrRh$_{2}$B$_{2}$C, consistent with the theory of White and Fulde \cite{8}, the $\gamma$-value being inversely proportional to the separation between the ground state and the excited state. For a better insight of excitonic mass enhancement in Pr-compounds with nonmagnetic singlet ground state we present the values of Sommerfeld coefficient $ \gamma $, crystal field splitting energy $ \Delta $ and low temperature saturation value of magnetic susceptibility $\chi$ in Table I. A comparison of $ \gamma $ and $ \Delta $ in these systems clearly reflect a systematic variation in their value, a smaller $ \Delta $ results in an increased $ \gamma $ and vice-versa. All these results establish the more general nature of the mechanism of excitonic mass enhancement in Pr-compouds, and the mass enhancement due to low lying crystal field excitations can even lead to the development of heavy fermion state as in our system PrRh$_{2}$B$_{2}$C.

\begin{table}
\caption{\label{tab:table1} The Sommerfeld coefficient $ \gamma $, crystal field splitting energy $ \Delta $ and low temperature saturation value of magnetic susceptibility $\chi$ for few Pr-compounds having nonmagnetic singlet as ground state.}
\begin{ruledtabular}
\begin{tabular}{lccccc}
 Compounds & Ground & First & Splitting  & $ \gamma $ (mJ/ & $ \chi_0$ ($\times$ 10$^{-2}$ \\
 {} & State & Excited & Energy & mole K$^2$) & emu/mole)\\
 {} & {} & State & $\Delta$ (K) & {} & at 2 K\\
\hline
PrRh$_{2}$B$_{2}$C & singlet & singlet & 14 & 250 & 20.4 \\
Pr$_{2}$Rh$_{3}$Ge$_{5}$& singlet & singlet & 12 & 81 & 10.6 \cite{10} \\
PrPd$_{2}$B$_{2}$C & singlet & singlet & 50 & 16 & 3.8 \cite{14} \\
PrRhGe$_{3}$ & singlet & doublet & 70 & 10 & 3.8 \cite{15}\\
\end{tabular}
\end{ruledtabular}
\end{table}

\begin{figure}
\includegraphics[width=8.5cm, keepaspectratio]{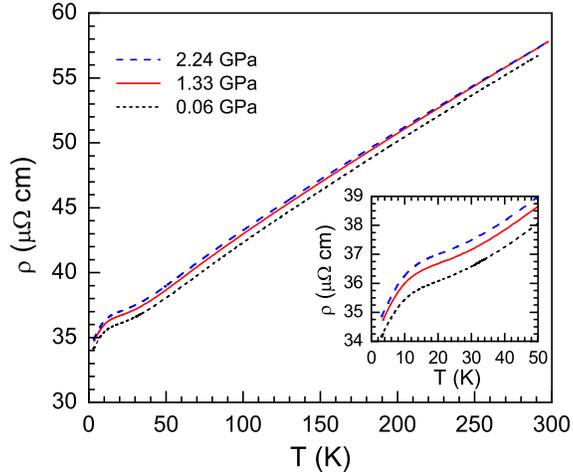}
\caption{\label{fig5} (colour online) Electrical resistivity, $\rho$(T), of PrRh$_{2}$B$_{2}$C measured under the application of external pressures up to 23 kbar. Inset shows the extended view of low temperature resistivity.}
\end{figure}

The application of pressure is expected to change the CEF level splitting energy and therefore to influence the physical properties. We therefore made an effort to achieve a stable ordered phase in PrRh$_{2}$B$_{2}$C by applying external pressure. In the resistivity studies shown in Fig. 5 no significant effect (except a slight increase in magnitude of resistivity) of pressure was observed up to 23 kbar. Especially, the position of the anomaly around 10 K attributed to the CEF effect is independent of pressure (the inflection point in $d \rho/dT$ characterizing the position of anomaly changes almost negligibly for the pressures of 0, 1.33 and 2.24 GPa) indicating that the applied pressure of up to 23 kbar has no considerable effect on the CEF level splitting scheme. Furthermore, this pressure is not sufficient to stabilize any superconducting or magnetically ordered phase in PrRh$_{2}$B$_{2}$C. The insensitivity to external pressure can be explained in the very rigid framework of the RNi$_{2}$B$_{2}$C structure as evidenced from a very high value of bulk modulus (e.g. YNi$_{2}$B$_{2}$C \cite{16}).

\section*{Conclusion}

We present clear evidence of heavy fermion behaviour in the new quaternary borocarbide PrRh$_{2}$B$_{2}$C. The mechanism for the electronic mass enhancement in this case is not the usual Kondo effect but it is due to the low-lying crystal field excitations. In this compound the ground state is a singlet separated from the first excited state only by about 10 K. Our effort to induce magnetism or superconductivity using hydrostatic pressure did not succeed. This is attributed to extremely rigid frame of the borocarbide structure as evidenced from high bulk modulus of YNi$_2$B$_2$C.

\section*{Acknowledgement}

We acknowledge CSR Indore (India) for providing access to low temperature measurements using PPMS.

\end{document}